\documentclass[preprint,showpacs,showkeys,floatfix]{revtex4}
\usepackage{epsfig}
\begin{document}

\title{Scintillating double beta decay bolometers}

\author{\firstname{S.}~\surname{Pirro}}\email{Stefano.Pirro@mib.infn.it}

\affiliation{Dipartimento di Fisica dell'Universit\`{a} di Milano-Bicocca and INFN, Sezione di 
                 Milano, I-20126 Milano, Italy}

\author{\firstname{J.W.}~\surname{Beeman}}
\affiliation{Lawrence Berkeley National Laboratory , Berkeley, California 94720, USA}

\author{\firstname{S.}~\surname{Capelli}}
\affiliation{Dipartimento di Fisica dell'Universit\`{a} di Milano-Bicocca and INFN, Sezione di 
                 Milano, I-20126 Milano, Italy}

\author{\firstname{M.}~\surname{Pavan}}
\affiliation{Dipartimento di Fisica dell'Universit\`{a} di Milano-Bicocca and INFN, Sezione di 
                 Milano, I-20126 Milano, Italy}

\author{\firstname{E.}~\surname{Previtali}}
\affiliation{Dipartimento di Fisica dell'Universit\`{a} di Milano-Bicocca and INFN, Sezione di 
                 Milano, I-20126 Milano, Italy}

\author{\firstname{P.}~\surname{Gorla}}
\affiliation{Laboratori Nazionali del Gran Sasso, I-67010, Assergi (L'Aquila), Italy}

\vskip 0.4cm

\begin{abstract}
We present the results obtained in the development of scintillating Double Beta Decay bolometers.
Several Mo and Cd based crystals were tested with the bolometric technique. The scintillation
light was measured through a second independent bolometer.
A 140 g CdWO$_4$ crystal was run in a 417 h live time measurement. Thanks to the scintillation
light, the $\alpha$ background is easily discriminated resulting in \emph{zero} counts above the 2615 keV
$\gamma$ line of $^{208}$Tl.
These  results, combined with an extremely  easy light detector operation, represent the first tangible proof 
demonstrating the feasibility of this kind of technique.
\end{abstract}


\maketitle

\section{Introduction}
The evidence of a neutrino rest mass represents one of the most exciting 
discoveries in the field of particle physics. 
The discovery of the neutrinoless Double Beta Decay (0$\nu$-DBD), however, will provide not only the 
ultimate answer about
the nature (Dirac or Majorana) of the neutrino, but will also allow to increase the sensitivity 
on the neutrino mass  down to a few meV.
As pointed out very recently by the Members of the APS Multidivisional Neutrino Study \cite{apsstudy},
Double Beta Decay searches will play a central role in the neutrino physics of the next decade.
The use of the bolometric technique offers the unique  possibility to investigate different DBD nuclei with 
a considerably higher energy  resolution, as needed for future experiments. 
In the case of a scintillating bolometer, the double independent read-out (heat and scintillation) will 
allow, thanks to the different scintillation Quenching Factor (QF) between $\alpha$ and $\gamma$, the 
suppression of the background events due to degraded $\alpha$-particles, the main source of background for
bolometric  0$\nu$-DBD experiments \cite{CUORE-PROPOSAL}.

\section{Environmental Background}
The experimental signature of the 0$\nu$-DBD is in principle very clear: 
a peak (at the $Q_{\beta\beta}$ value) in the two-electron summed energy spectrum.
In spite of this characteristic imprint, the rarity of the process  makes the identification very difficult.
Such  signals have to be disentangled from a background due to natural radioactive decay chains,
cosmogenic-induced activity, and man-made radioactivity, which deposit energy in the same region 
as the DBD,  but at a faster rate. 
Consequently, the \emph{main task} in 0$\nu$-DBD searches is the natural background suppression  
using the state-of-the-art ultra-low background techniques and, hopefully, identifying the signal.
There are different sources of background for DBD experiments that can be classified in four main categories.

\subsection{External $\gamma$  background}
\begin{figure}
\begin{center}
\includegraphics[width=0.8\textwidth]{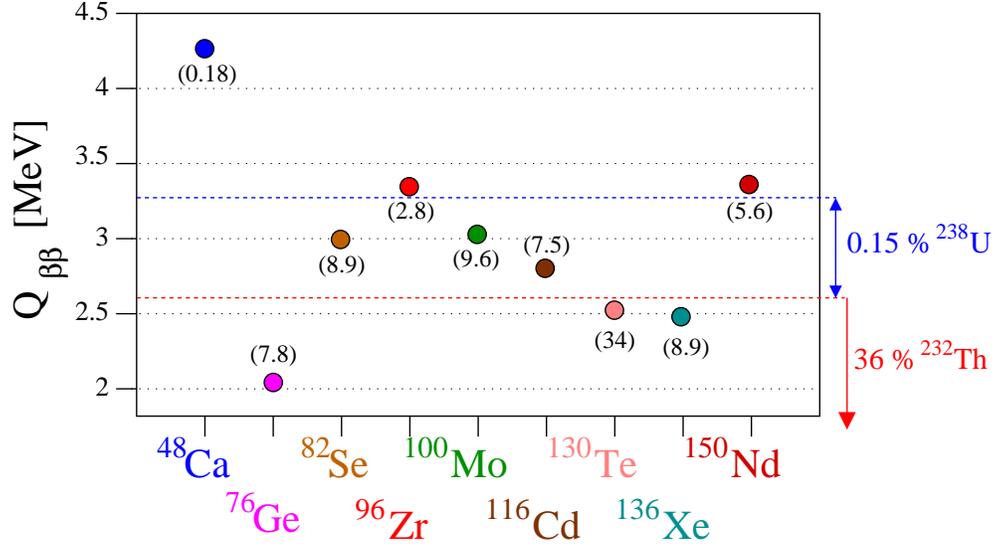}
\caption{\label{fig:fig1}Most interesting DBD emitters nuclei. In parenthesis the natural isotopic abundance.}
\end{center}
\end{figure}

Environmental $\gamma$'s represent the main source of background for most of the present 
DBD experiments and arise mainly from the  natural  contaminations in $^{238}$U and  $^{232}$Th. 
The  common highest $\gamma$ line  is the 2615 keV line of  $^{208}$Tl, with a total branching 
ratio (BR) of 36 \% in the $^{232}$Th decay chain.
Above this energy there are only  \emph{extremely rare} high energy $\gamma$-rays from  $^{214}$Bi; 
the total BR in the energy window from 2615 up to 3270 keV is 0.15 \% in the $^{238}$U decay chain.
It is therefore clear that a detector based on an DBD-emitter with the $Q_{\beta\beta}$ value above the
2615 keV line of  $^{208}$Tl represents the optimal starting point for a future experiment.
The most interesting DBD nuclei are shown  in Fig.~\ref{fig:fig1}.

\subsection{Neutrons}

Another important source of background is due to neutrons: low energy neutrons can induce (n,$\gamma$) reactions
in materials close  (or internal) to the detectors with $\gamma$ energies up to 10 MeV; 
furthermore, high energy neutrons generated by $\mu$-induced spallation reactions can release 
several MeV  by direct interaction in the detectors. 
This contribution represents only  1\% -- 10\% of the total background for the \emph{present} DBD experiments.
Unlike $^{238}$U and  $^{232}$Th trace contaminations, neutrons can be, at least in principle, suppressed 
with suitable use of shielding/veto's. 

\subsection{Surface contaminations}
This source of background plays a role  for almost all detectors, but is crucial for
\emph{fully active} detectors, as in the case of bolometers.

\begin{figure}
\begin{center}
\includegraphics[width=0.6\textwidth]{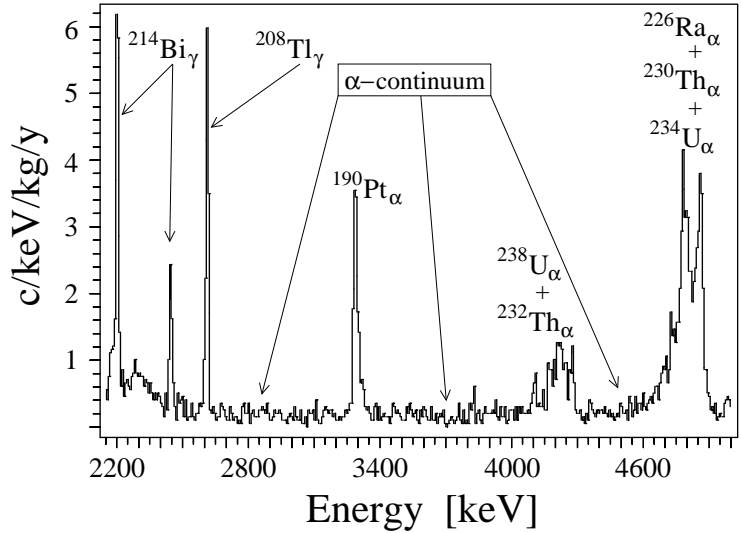}
\caption{\label{fig:fig2}Background spectrum of the CUORICINO experiment. The $\alpha$-continuum 
extends up to the 0$\nu$-DBD region, at 2528 keV.}
\end{center}
\end{figure}

$\alpha's$ arising from surface contaminations located in dead layers faced to (or on) the detectors can lose part 
of their energy in a few microns and reach the detectors with an energy corresponding to the  $Q_{\beta\beta}$ value.
It is not straightforward to deal with this problem due to the fact that, unlike bulk contamination 
(that can be measured through HPGE detectors), the sensitivity of standard diagnostic devices can hardly  
reach the needed sensitivity \cite{Pirro-LRT-2005}.
These contaminations arise from the machining, cleaning, or passivation of the materials to be used close
to the detectors, as well from Radon implantation. This $\alpha$-continuum represents the main source of background
for the CUORICINO DBD experiment \cite{CUORICINO}, as  can be deduced from Fig.~\ref{fig:fig2}.

\subsection{$^{238}$U and  $^{232}$Th internal contaminations}
\begin{figure}
\begin{center}
\includegraphics[width=0.8\textwidth]{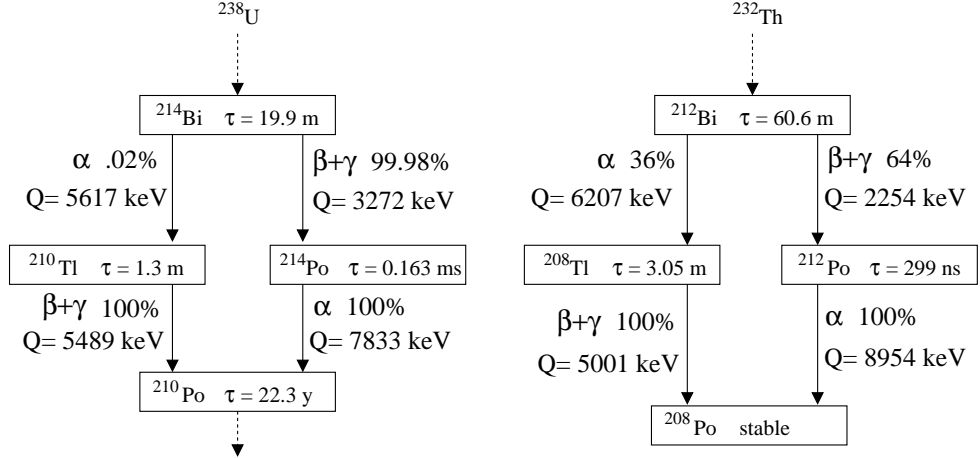}
\caption{\label{fig:fig3}High energy $\beta-\gamma$ decays of the  $^{238}$U and  $^{232}$Th chains that 
can mimic the 2 electron signal.}
\end{center}
\end{figure}

This source of background has to be considered very carefully for non-homogeneous (or passive) detectors, but
under certain assumptions does not play
 a significant role for  homogeneous detectors.
In the case of homogeneous detectors, in fact, dangerous $\beta-\gamma$ events can be recognized through delayed 
$\alpha$ coincidences.  
If we discard the contribution of $^{234}$Pa of the $^{238}$U chain, that has  $\beta-\gamma$ events 
with Q$_{tot}$=2195 keV, all the remaining high energy decays are shown in Fig.~\ref{fig:fig3}.
As can be argued by the scheme, the  $\beta-\gamma$ decays are preceded  (or followed) by an $\alpha$ emission.
Therefore, using delayed  $\alpha$ coincidences, $\beta-\gamma$ decays, that can mimic the 2 electron signal,
are discarded. This technique can be easily applied  for the $^{238}$U decay chain, while may have some problems
(large death time) with the $^{208}$Tl decay. In this case the decay is preceded by the $\alpha$ of 
$^{212}$Bi with a mean time given by T$_{1/2}$=3.05 m, and it is therefore clear that this method holds only 
if the contaminations are not too large.

\subsection{Cosmogenic activity}
This source of background can affect both the detector itself and the surrounding shielding materials. 
Regarding the detectors, the production of long-lived radioactive nuclei depends crucially on the target material. 
Probably the best known cosmogenic isotope is the $^{60}$Co. This nucleus  is extremely 
dangerous both for internal (detector) and for external (shielding) contaminations. 
$^{60}$Co is, unfortunately, a common contaminant of Copper, 
which represents the cleanest solid material available so far, often used for internal radioactive 
shielding by several DBD experiments. 
With respect to  internal contamination, the background spectrum is due to the 
$\beta$ (Q=318 keV) + 2$\gamma$ (1173+1332 keV), so that the  released energy can reach the total Q-value (2824 keV).
Regarding external contaminations, the background is mostly due  only to the 2 $\gamma$'s (emitted in coincidence), 
with total energy of 2505 keV. 
Therefore a future experiment based on a nucleus with $Q_{\beta\beta} >$ 2505 (and possibly also $>$2824 keV) will
avoid this problem.

\section{Bolometric light detectors}   
\label{sec:ld}
The first idea of using a scintillating bolometer was suggested for solar neutrino experiments 
in 1989 \cite{Gallix89}.
The first light/heat measurement with $\alpha$ background discrimination  for DBD searches was performed 
with a thermal bolometer and  silicon photodiode by our group in 1992 \cite{Ale92}, but was  no more  pursued 
due to the difficulties of running such a light detector at low temperatures ($\simeq$ 10 mK). 
The idea to use  a bolometer as light detector  was first developed by C. Bobin et al.  \cite{Coron97} and 
then later optimized  \cite{CRESST,ROSEBUD}, for Dark Matter  searches. 
Starting from that work we developed a thermal light detector (LD) to be used  for DBD search. 

\begin{figure}
\begin{center}
\includegraphics[width=0.5\textwidth]{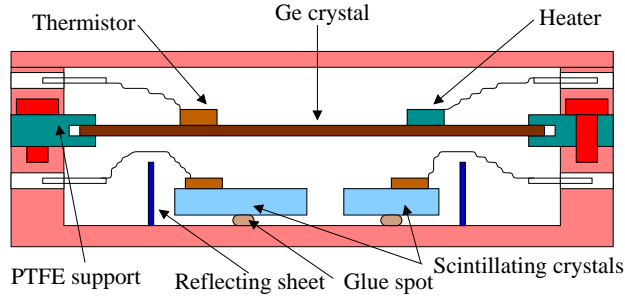}
\caption{\label{fig:fig4}LD + scitillating crystals setup.}
\end{center}
\end{figure}

A bolometer is, in principle, a very simple device. It is composed of an absorber coupled with a 
thermometer, so that an energy release in the absorber can be detected.
The temperature rise in a thermal bolometer is
given by $\Delta T \propto E/C$, where E represents the energy released into the detector and C 
its heat capacity. This means that a  very small detector can reach a very high sensitivity (few tenths of eV);
therefore a ``dark'' thin bolometer can absorb scintillation photons and give  a measurable thermal signal. 
I our case, the bolometer serves primally as a light detector and  has the characteristic time constant 
of bolometers (20--500 ms). 
However it is also sensitive  to every energy release ( $\alpha's,\beta's,\gamma's$) and acts
as a particle detector.
Certainly large-surface bolometric LD's cannot easily reach the threshold of PMT's 
($\sim$ 1 photoelectron i.e.  3--7 photons, tacking into account the quantum efficiency conversion), but 
they have two important advantages: (i)  they are sensitive over an extremely large band of photon 
wavelength (depending on the absorber); (ii) the overall quantum  efficiency can be as good as that of photodiodes. 
This means that the energy resolution on the scintillating
light, which depends (above  threshold)  only on the Poisson statistical fluctuation of the emitted photons, will be
better for bolometric LD's with respect to PMT's (see Sec.~\ref{sec:cdwo4}).
The main characteristics of a bolometric light detector should be  the \emph{easy} expandability up to 
$\sim$1000 channels, and the \emph{complete reliability} of the composed device (bolometer + LD) 
in order to have an almost 100\% live time measurement. 
On the other hand, there is not the need to  have an extremely sensitive detector, since the 
DBD signal lies in the MeV range. \\
We developed our first LD as a pure Ge disk absorber (35 mm diameter, 1 mm thick) thermally coupled with an
3x1.5x0.4 mm$^3$ Neutron Transmutation Doped  Ge thermistor (thermometer).
We adopted  a very simple setup in which the  disk is held by two PTFE supports squeezed by a screw at two 
opposite sides. The complete setup (see Fig.~\ref{fig:fig4}) was tested in the CUORE \cite{CUORE} R\&D cryostat 
located deep underground in the Gran Sasso National Laboratories (Italy). The description of the 
Front-end electronic can be found in \cite{precaldo,prefreddo}. 

The LD was used to test several scintillating crystals (see Sec.~\ref{sec:molibdates}).
After the encouraging results we decided to optimize the light detector.

At the working point of our bolometer, $\sim$ 13 mK, the theoretical heat capacity  of the Ge absorber  is
$\sim$ 0.2 pJ/K while that of the thermistor is $\sim$ 32 pJ/K. Therefore the signal height is 
limited practically  by the thermistor itself. 

We therefore decided to increase the size of the Ge absorber.
The idea was to develop a large-area light detector, able to read several \emph{large} scintillating 
crystals at the same time, lowering the number of LD's needed to face a very large detector array.
We used a 6.6 cm diameter, 1 mm thick Ge crystal (absorber) coupled with  the same kind 
of thermistor used for the previous LD.
Furthermore   one side of the crystal (the one facing the scintillator) was coated with a layer of 60 nm of SiO$_2$ 
in order to increase the absorption of the scintillating photons. 
The new LD was mounted in the same way as the previous one. 
We tested the new large-area LD with a  3x3x2 cm$^3$, 140 g CdWO$_4$ single crystal (see Sec.~\ref{sec:cdwo4}). 

\section{Results on Molybdates}
\label{sec:molibdates}
\begin{figure}
\begin{center}
\includegraphics[width=0.8\textwidth]{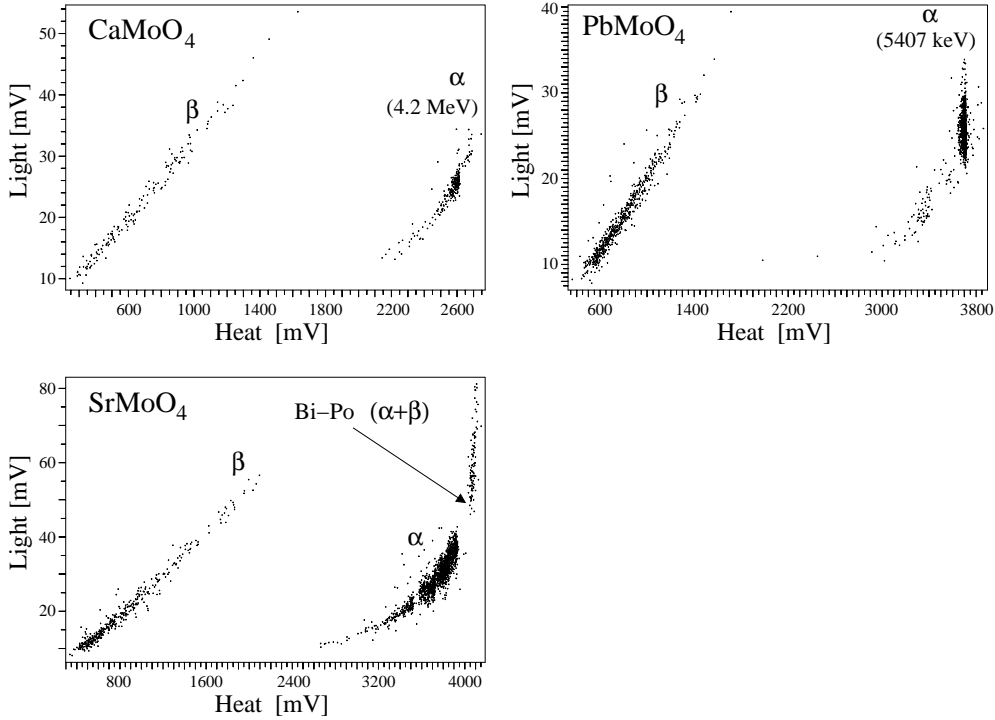}
\caption{\label{fig:fig5}Scatter plots of Light vs Heat obtained with the three crystals. 
All the crystals, due to their small size, saturate the heat channel at high energies. The Heat energy 
scale is therefore not linearized and expressed in mV. The Y axis represents the light output (in mV) of the LD. 
In this case the scale is linear and direct comparison between the three crystals is possible.}
\end{center}
\end{figure}

Within the \emph{standard} scintillators, an obvious possibility for DBD searches are the Molybdates, due to the high
$Q_{\beta\beta}$ of $^{100}$Mo (3034 keV).
The scintillation properties of these  crystals, based on the (MoO$_4$)$^{2-}$ oxyanions, were studied only 
very recently \cite{Rodny-2002}, and further investigations are on the way.
Our first aim was to see the scintillation light yield (LY) of several different samples.
At that time no direct measurements were available on the particle-induced LY at temperatures lower than 77 K. 
Our  LD was faced (simultaneously) to  several small size (few grams)  molybdate crystal bolometers  
(PbMoO$_4$, CaMoO$_4$, SrMoO$_4$) and all  of them exhibited scintillation light (see Fig.~\ref{fig:fig5}).
Due to the small size of the molybdate crystals, their energy response, as a bolometer, becomes extremely 
non-linear at high values. This behaviour is common for small bolometers and can be corrected 
using calibration sources. 
In our case the only calibration source used was a pure  $^{238}$U solution evaporated on a small spot facing each 
crystal. The only ``clear'' peak is therefore the double $\alpha$ line of $^{238}$U at 4198 and 4151 keV. 
Besides this $\alpha$ decay there is the $\beta$ decay of $^{234}$Pa with an endpoint at 2.3 MeV. 
The induced scintillation due to the $\beta$ + 2 $\gamma$ decays of $^{234}$Th was below the threshold of the LD.

The three scatter-plots in Fig.~\ref{fig:fig5} show some differences, particularly in the $\alpha$ region.
Regarding the CaMoO$_4$ sample the $^{238}$U is clearly visible in the scatter plot.
In the case of the PbMoO$_4$ sample, the scatter plot is dominated by an internal contamination of $^{210}$Pb,
with the characteristic   $\alpha$ line of $^{210}$Po at 5407 keV.
The situation is not completely clear in the SrMoO$_4$ sample: this crystal shows extremely large contaminations 
in the $\alpha$-region and   is very difficult to disentangle them due to the extreme non-linearity of the detector. 
Furthermore, a clear Bi-Po event ($\alpha+\beta$) is present in the spectrum (due to the slowness of the thermal
detectors the two decays appear as one). 
Even if   not exactly quantifiable due to the non-linearity of the energy scale, the power of  
the $\alpha$ discrimination technique is absolutely clear in Fig.~\ref{fig:fig5} thanks to the different 
scintillation QF.

The second step could be to use larger crystals. Once we have experimentally proven that the LY is sufficient 
to discriminate the $\alpha$'s, we also have to show that we can  build  large size 
(hundreds of grams) bolometers with the needed high energy resolution. 
However other considerations need to be addressed first in order to have a good scintillating 
bolometer for a future high sensitivity DBD experiment.

PbMoO$_4$ has two disadvantages: it contains Lead and therefore  the radioactive $^{210}$Pb. 
The only way to avoid this problem would be to use ancient lead to grow the crystal. 
Furthermore the mass fraction of Molybdenum  within the compound will be  ``only'' 26 \%.

Our sample of SrMoO$_4$ seems  to be  radioactively contaminated, so we can not fully judge this material. 
With some effort, a radiopure crystal might be grown but the 
cosmogenic $^{90}$Sr, in spite of its low  $Q_{\beta}$ energy,  could represent a problem, especially 
regarding pile-up.

CaMoO$_4$ could be a good candidate. It contains a large fraction of Molybdenum, there are several producers 
able to grow large crystals, and it seems to be reasonably radiopure. Unfortunately the backdraw is  Ca.
The 2-neutrino DBD of $^{48}$Ca, with  $Q_{\beta\beta}$=4271 keV  (although only 0.18 \% of natural isotopic 
abundance) will result in an unavoidable  background in the 0$\nu$-DBD region of $^{100}$Mo. This background
can be easily evaluated as 0.01 c/keV/kg/y.

\section{Experimental results on CdWO$_4$} 
\label{sec:cdwo4}
\begin{figure}
\begin{center}
\includegraphics[width=0.8\textwidth]{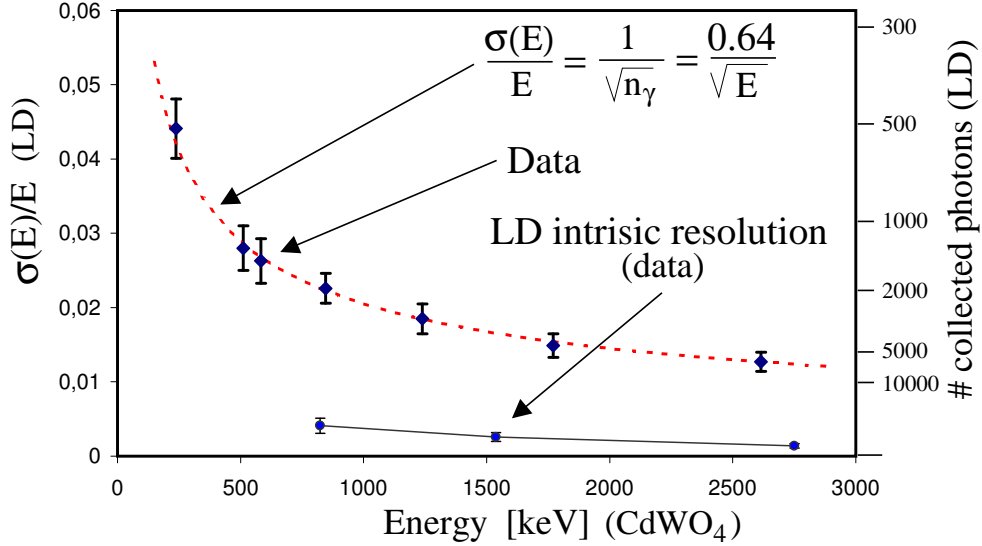}
\caption{\label{fig:fig6}Energy resolution of the Light Detector on the scintillation light 
emitted by the CdWO$_4$ corresponding to the full energy peaks.
The intrinsic energy resolution is evaluated thanks to the heater pulses.}
\end{center}
\end{figure}

This bolometer  was tested, together with 8 CUORE bolometers, in a long background measurement during April 2005.
The 3x3x2 cm$^3$, 140 g CdWO$_4$ single crystal was almost fully surrounded by a reflecting sheet 
oriented towards the LD that and glued, with a small  spot, on the copper.
It was not possible to calibrate the  LD with the standard $\gamma$-sources we usually place \emph{outside}   
the cryostat, due to its small size  (it is almost transparent to $\gamma$'s, while external X-rays 
would not reach the cryostat inner volume). Nevertheless an estimation can be obtained 
by considering the resolution on the light signal. Assuming that the resolution of the LD on the scintillation 
peaks is  only due to the statistical fluctuation of the number of photons produced in the CdWO$_4$ 
crystal, the LD can be calibrated. This assumption holds only if the intrinsic energy resolution of the detector 
is small with respect to the photon fluctuation. The intrinsic energy resolution of the LD can be evaluated
by means of a heater that can inject monochromatic thermal energy ~\cite{ALES98} into the detector.
We found in this way that the intrinsic energy resolution  was negligible with  respect to 
the width of the  scintillation peaks. 
Therefore, assuming a Fano factor equal to 1, and plotting the energy of the photopeaks vs the energy resolution, 
an energy calibration (in photons) can be found as shown in  Fig.~\ref{fig:fig6}.
The number of collected photons was measured to be  $\sim$ 2400 photons/MeV. 
Using this calibration, and assuming $\sim$ 3 eV/photon the LD can be calibrated in energy. 
With this method we found that the energy resolution of the LD is $\approx$~48 eV FWHM (or 16 photons). 
It turns out that the  baseline resolution of the LD was observed to be 1 order of magnitude smaller compared to 
the intrinsic statistical photon fluctuation at 2615 keV, suggesting the possibility to further increase 
the area of the LD for our purposes.

The  energy resolution on the scintillation peak  at 2615 keV is 3\% FWHM. 
This value is similar (in some cases even better) to those obtained with NaI detectors.
Furthermore this  result is much better (8\% FWHM) with respect to the one  obtained with \emph{dedicated}
DBD experiments  using CdWO$_4$ as a scintillator (at room temperature) read by standard PMT's ~\cite{Danevich2003}.
A direct comparison between the two set-ups is, however, not straightforward.
First of all the size of the crystals is different (330 g in that case). Furthermore there could be a LY increase 
in the CdWO$_4$ with the temperature decrease, therefore justifying our better result.
But probably the explanation of our better energy resolution  is due to the quantum efficiency, as
explained in Sec.~\ref{sec:ld}: the PMT's matched for CdWO$_4$ emission have a quantum efficiency 
of the order of only 18\%. 
Furthermore the best energy resolution ever  achieved with PMT's on CdWO$_4$ scintillators  
is 3.7 \%  FWHM (at 2615 keV) ~\cite{Danevich2004} on  an extremely small sample (1 cm$^3$). 

\begin{figure}
\begin{center}
\includegraphics[width=0.67\textwidth]{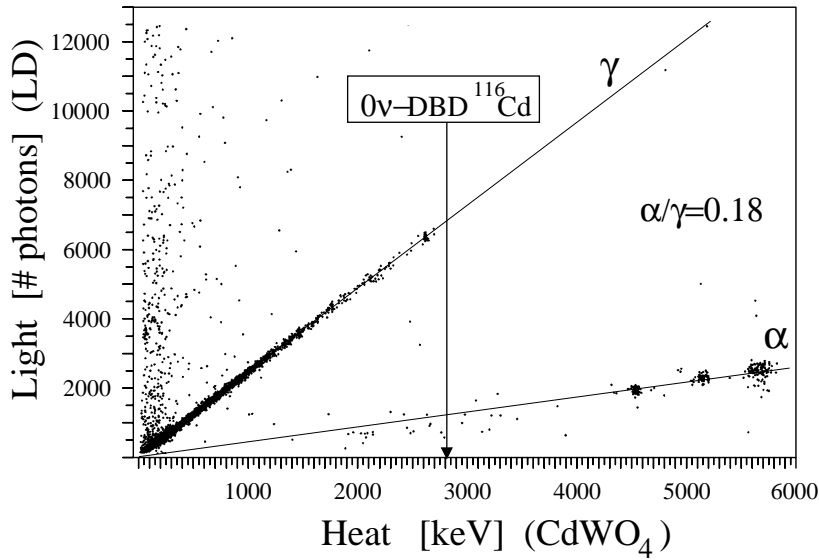}
\caption{\label{fig:fig7}Scatter plot of Heat vs Light obtained in a  417 h live time background measurement. 
The points in the 0-300 keV (Heat) region are due to pile-up. 
The $\alpha$-induced background is clearly discriminated.}
\end{center}
\end{figure}

We also performed a long background run  with the CdWO$_4$-LD  device. Since the main aim of the run was to 
perform a low background measurement with 8 CUORE detectors, the device was mounted ``far away'' from
the TeO$_2$ detectors. Because of  the limited experimental volume in the cryostat,  the 
device was mounted only partially shielded (not shielded against the dilution units itself) and not in 
the ``best'' position (the CUORE detectors are connected to a vibration decoupler that  minimizes the mechanical 
vibrations that reach the detectors causing thermal noise).
During the 417 h live time background measurement the energy resolution of the CdWO$_4$ was 25 keV.
The obtained scatter plot Heat vs Light of our detector is shown in Fig.~\ref{fig:fig7}.
A live time of 80\% was achieved.

The $\alpha$-continuum is completely ruled out thanks to the scintillation detection. Moreover the background 
above the 2615 keV line  is practically \emph{zero}, demonstrating the power of this technique.
Only two counts are present, above the $^{208}$Tl line, around $\sim$ 5 MeV. 
One of these seems to be a pile-up (it is just above an $\alpha$ line), while  the second  
could arise from the (n,$\gamma$) reaction on $^{113}$Cd ($\sigma \simeq$ 20000 barn).
Our cryostat is not shielded against thermal neutrons and tacking into account the natural 
neutron flux in our laboratory, we expect $\sim$ 0.8 $\gamma$ counts in the energy window 2.6 - 9 MeV.

The large number of events in the 0--300 keV region is 
due to the natural $\beta$ decay of $^{113}$Cd (12.2 \% i.a.) with Q$_\beta$=318 keV 
and T$_{1/2}$=7.7 10$^{15}$ y. The expected activity can therefore be evaluated (and directly measured) and
turns out to be 0.12 Hz. 

Some pile-up is found in the scatter plot, especially clear in the 0-300 keV ``Heat'' region.
This was due to the fact that ``close'' to the LD (and therefore close to the CdWO$_4$) a 
relatively intense $^{238}$U source was present in order to calibrate other bolometers mounted in the 
cryostat in this run. Since the LD is a bolometer, it exhibits the same pulses whether it 
absorbs photons or particles.
Due to the slowness of the LD some pile-up  appears in the scatter plot. 
This pile-up can be reduced  with a combined trigger (in our case the 
two detectors had independent triggers). In any case the problem can be avoided simply eliminating the 
$^{238}$U source. 

\begin{figure}
\begin{center}
\includegraphics[width=0.67\textwidth]{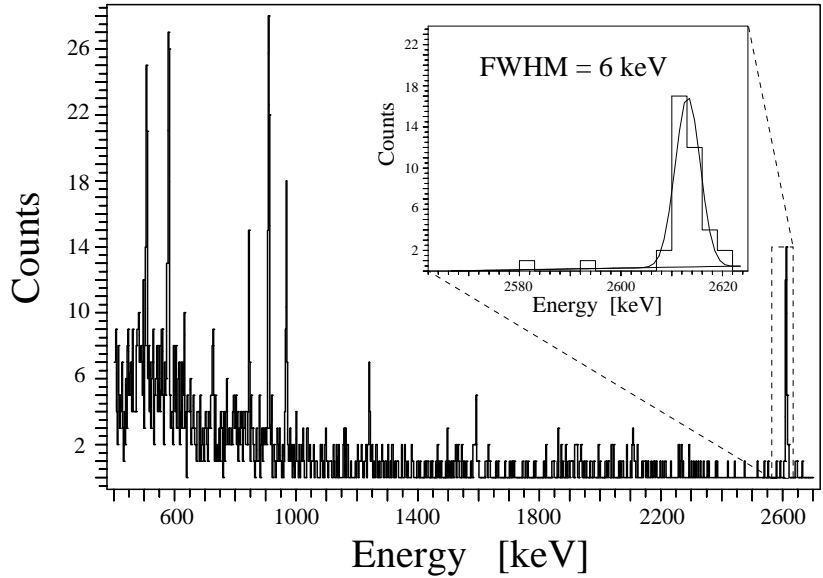}
\end{center}
\caption{\label{fig:fig8}Calibration spectrum of the CdWO$_4$ detector exposed to an external source of $^{232}$Th.
In the inset is shown the energy resolution on the 2615 keV line.}
\end{figure}

At the end of the background measurement we recognized a small failure in the electronic amplifier
board of the CdWO$_4$. We fixed the problem and we performed a new energy calibration using a $^{232}$Th source.
The obtained calibration spectrum is shown in Fig.~\ref{fig:fig8}.

The energy resolution at the 2615 keV line is 6 keV FWHM, which represents the typical value 
that can be reached with bolometers.

In the above discussion, we have shown that light measurement helps to identify the $\alpha$-induced background
in thermal detector schemes. However, this technique
is also extremely helpful for rejecting another ``unavoidable'' source of background that can take
place  with thermal detectors.
It was observed \cite{CRESST2} that rare heat releases induced by materials relaxations can induce thermal
pulses indistinguishable with  respect to particle-induced thermal pulses. It is presently not completely 
clear the energy scale of such events, but in any case a \emph{completely  independent} double read-out of the 
signal (heat and scintillation) will also allow one to identify such events due to the fact that they do not produce
scintillation light.

\section{Conclusions}
We developed a large area (34 cm$^2$) thermal light detector for DBD searches. It is by far the
largest bolometric light detector ever operated. The device is characterized by  
an easy construction and assembly  and by an intrinsic, constant,  energy resolution of the order 
of 16 photons FWHM. The results obtained are competitive with standard NaI scintillators.
We tested several DBD scintillating bolometers, demonstrating the feasibility of such a technique.

For the first time a DBD \emph{pilot experiment} based on simultaneous detection of heat and light 
was performed on a large DBD scintillating crystal   showing \emph{directly} the feasibility 
and the reliability of this technique. 
The background  that can be obtained with such  detectors,
provided that the Q$_{\beta\beta}$ is above 2615 keV, can \emph{easily} reach levels at least 3 orders
of magnitude better with respect to the present experiments. Furthermore the bolometric
technique is the only one that allows one to investigate most of the DBD emitters with the required 
energy resolution.
In the  future we plan to test  other molybdate crystals as well as different interesting compounds
(ZnSe, ZrO$_2$).    

\section{Acknowledgments}
We are particularly gratefully to Prof. P.A. Rodnyi of the St. Petersburg State Technical University for
providing (as a gift) three small molybdate crystals.
We also acknowledge Dr. Vitalii Mikhailik of the Physics Department of  the Oxford University for  useful 
discussions and collaborations.

\end{document}